\documentclass{ws-procs975x65}

\begin{document}
%to switch ON running title
%\markboth{J. Madsen}{Strangelets in Cosmic Rays}

%\wstoc{Strangelets in Cosmic Rays}{J. Madsen}

\title{ Strangelets in Cosmic Rays }
\author{ Jes Madsen} 
\address{Department of Physics and Astronomy, University of Aarhus,\\
DK-8000 {\AA }rhus C, Denmark\\
\email{jesm@phys.au.dk}} 

\begin{abstract}
The properties of strangelets are reviewed and two experiments
searching for them in cosmic rays are described. The prospects for
strangelets as ultra-high energy cosmic rays beyond the classical
GZK-cutoff are discussed.
\end{abstract}

\bodymatter

\section{Bulk quark matter%
\protect\footnote{Sections 1 and 2 are partly based on
Ref.~\protect\citen{Madsen:1998uh}.}
}

Quark matter composed of $u$ and $d$-quarks is
unstable except for 3-quark baryons (otherwise nuclei would decay to quark
matter). An additional Fermi sea with a third flavor, $s$, makes
it possible to reduce the energy by
increasing the spatial density of quarks. The $s$-quark mass is high
compared to that of $u$ and $d$, so stability is most likely for low 
$s$-quark mass
\cite{Bodmer:1971we,Chin:1979yb,Witten:1984rs,Farhi:1984qu}.

The Fermi momentum, $p_{Fi}$, of a non-interacting, massless quark-flavor, $i$,
equals the chemical potential, $\mu _{i}$. The number density is $%
n_{i}=\mu _{i}^{3}/\pi ^{2}$, the energy density $\epsilon _{i}=3\mu
_{i}^{4}/(4\pi ^{2})$, and the pressure $P_{i}=\mu _{i}^{4}/(4\pi ^{2})$.
The sum of the quark pressures is balanced by the confining bag pressure, $B$%
; $\sum_{i}P_{i}=B$; the total energy density is $\epsilon =\sum_{i}\epsilon
_{i}+B=3\sum_{i}P_{i}+B=4B$, and the baryon number density is $%
n_{B}=\sum_{i}n_{i}/3$. For a gas of $u$ and $d$-quarks charge neutrality
requires $n_{d}=2n_{u}$, or $\mu _{2}\equiv \mu _{u}=2^{-1/3}\mu _{d}$. The
corresponding two-flavor quark pressure is $P_{2}=P_{u}+P_{d}=(1+2^{4/3})\mu
_{2}^{4}/(4\pi ^{2})=B$, the total energy density $\epsilon _{2}=3P_{2}+B=4B$%
, and the baryon number density $n_{B2}=(n_{u}+n_{d})/3=\mu _{2}^{3}/\pi
^{2} $, giving an energy per baryon of 
\begin{equation}
\epsilon _{2}/n_{B2}=(1+2^{4/3})^{3/4}(4\pi
^{2})^{1/4}B^{1/4}=6.441B^{1/4}\approx 934{\rm MeV}B_{145}^{1/4}, 
\end{equation}
where $B_{145}^{1/4}\equiv B^{1/4}/145{\rm MeV}$; 145MeV being the lowest
possible choice that avoids instability of ordinary nuclei.

A three-flavor quark gas is electrically neutral for $n_{u}=n_{d}=n_{s}$,
i.\ e.\ $\mu _{3}\equiv \mu _{u}=\mu _{d}=\mu _{s}$. For fixed bag constant
the three-quark gas should exert the same pressure as the two-quark gas
(leaving also the energy density, $\epsilon _{3}=3P_{3}+B=4B$, unchanged).
That happens when $\mu _{3}=[(1+2^{4/3})/3]^{1/4}\mu _{2}$, giving a baryon
number density of $n_{B3}=\mu _{3}^{3}/\pi ^{2}=[(1+2^{4/3})/3]^{3/4}n_{B2}$%
. The energy per baryon is then 
\begin{equation}
\label{eps3}
\epsilon _{3}/n_{B3}=3\mu _{3}=3^{3/4}(4\pi
^{2})^{1/4}B^{1/4}=5.714B^{1/4}\approx 829{\rm MeV}B_{145}^{1/4}; 
\end{equation}
{\it lower\/} than in the two-quark case by a factor
$n_{B2}/n_{B3}=(3/(1+2^{4/3}))^{3/4}\approx 0.89$, so one may gain of order
100 MeV per baryon by introducing an extra flavor.

For massive $s$-quarks weak equilibrium gives $\mu_s=\mu_d=\mu_u+\mu_e$,
where $\mu_e>0$ corresponds to a small but non-zero density of electrons
required for local charge neutrality. For reasonable values of $m_s$ of
order 100~MeV stable strange quark matter remains possible for a range of
$B$.

Quark matter at asymptotically high density has an interesting property
called color superconductivity \cite{Alford:1997zt,Rapp:1997zu}. 
Even the weakest attraction (and such
attractions exist in QCD) leads quarks of different colors and flavors to
form pairs, much like Cooper pairs in a superconductor, except that the
binding in QCD is caused by a direct attraction channel rather than via
indirect phonon interaction. The binding energy of a pair, $\Delta $, can be
very large, ranging from a few MeV to over $100\,{\rm MeV}$. In general
these systems are called color superconductors. If all colors and flavors
pair in an equal manner one talks about color-flavor locking.

While color-flavor locking seems generic in the infinite
density limit, the properties of strange quark matter at densities of order
or somewhat higher than nuclear matter density is at the focus of much
current research and discussion. This is the density regime of relevance for
strangelets, strange stars, and for quark matter cores in hybrid stars (the
analogs of neutron stars if quark matter is metastable so that it forms
above a certain density in compact star interiors). An additional binding
energy per baryon of approximately $-3\Delta ^{2}/\mu $ is introduced in
these systems, meaning that extra binding of $10-100\,{\rm MeV}$ is not
unrealistic, thus significantly increasing the likelihood of absolutely
stable strange quark matter and strangelets.

\section{Strangelets}
\label{sec:strangelet}

For $A\ll 10^7$ the quark part of strange quark matter
is smaller than the Compton wavelength of electrons,
so electrons no longer ensure local charge neutrality. Therefore Coulomb
effects have to be included, though the fortuitous cancellation of $%
q_u+q_d+q_s=\frac 23-\frac 13-\frac 13=0$ means that Coulomb energy is much
less important for strangelets than for nuclei. In particular
strangelets do not fission. For $A<10^3$ other finite size effects 
such as surface tension and curvature have to be taken into account.

Several strangelet searches with relativistic heavy-ion collisions as well
as cosmic ray searches have been carried out, and others are planned for the
future \cite{Klingenberg:2001qs,Sandweiss:2004bu,Finch:2006pq}. 
Most of these searches are sensitive only to low $A$-values, so it
is important to know the properties of small lumps of strange quark matter
(strangelets).

Mode-filling for large numbers of quarks in a spherical MIT-bag
\cite{Chodos:1974je} was
performed for $ud$-systems in Ref.~\citen{vasgre86}, and
for strangelets in 
Refs.~\citen{Farhi:1984qu,greal88,takboy88,Gilson:1993zs,Madsen:1994vp,schgre97}.
All of these calculations were performed for strong fine-structure
constant $\alpha _{s}=0$.

The energy per baryon smoothly approaches the bulk limit for $A\rightarrow
\infty $, whereas the energy grows significantly for low $A$. For low $s$%
-quark mass shells are found for $A=6$ (3 colors and 2 spin orientations per
flavor), and less conspicuous ones for $A=18$, 24, 42, 54, 60, 84, 102 etc.
As $m_{s}$ increases it becomes more and more favorable to use $u$ and $d$
rather than $s$-quarks, and the \textquotedblleft magic
numbers\textquotedblright\ change; for instance the first closed shell is
seen for $A=4$ rather than 6.

Strangelet mode-filling calculations can be modified by inclusion of Coulomb
energy and zero-point fluctuation energy. The Coulomb energy is generally
small. The zero-point energy is normally included as a phenomenological term
of the form $-Z_{0}/R$, where fits to light hadron spectra indicate the
choice $Z_{0}=1.84$ \cite{Gilson:1993zs}.
Roughly half of this phenomenological term is due to
center-of-mass motion. The proper choice of $\alpha _{s}$ and $Z_{0}$ is
tricky. As discussed by Farhi and Jaffe \cite{Farhi:1984qu} the values
are intimately coupled to $B$ and $m_{s}$, and it is not obvious that values
deduced from bag model fits to ordinary hadrons are to be preferred in
the case of strangelets. This
uncertainty may have an important effect for $A<5$--$10$, but the zero-point
energy quickly becomes negligible for increasing $A$ as it decreases
like $A^{-1/3}$. It means, however, that it is difficult to
match strangelet calculations to experimental data concerning ordinary
hadrons or limits on the putative $A=2$ $H$-dibaryon.

Mode-filling calculations give the \textquotedblleft
correct\textquotedblright\ results as far as the model can be trusted, but
for many applications a global mass-formula analogous to the liquid drop
model for nuclei is of great use and also gives further physical insight.

Berger and Jaffe\cite{Berger:1986ps} 
made such a detailed analysis within the MIT bag
model. They included Coulomb corrections and surface tension effects
stemming from the depletion in the surface density of states due to the mass
of the strange quark. Both effects were treated as perturbations added to a
bulk solution with the surface contribution derived from a multiple
reflection expansion.
A self-consistent treatment including also the very important curvature
energy was given in Refs.~\citen{Madsen:1994vp,Madsen:1993ka,Madsen:1993iw}.

In the ideal Fermi-gas approximation the energy (mass) of a system composed of
quark flavors $i$ is given by 
\begin{equation}
E=\sum_i(\Omega_i+N_i\mu_i)+BV+E_{{\rm Coul}}.  \label{Estrangelet2}
\end{equation}
Here $\Omega_i$, $N_i$ and $\mu_i$ denote thermodynamic potentials, total
number of quarks, and chemical potentials, respectively. $B$ is the bag
constant, $V$ is the bag volume, and $E_{{\rm Coul}}$ is the Coulomb energy.

In the multiple reflection expansion framework of Balian and Bloch \cite%
{balblo70}, the thermodynamical quantities can be derived from a density of
states of the form 
\begin{equation}
{\frac{{dN_{i}}}{{dk}}}=6\left\{ {\frac{{k^{2}V}}{{2\pi ^{2}}}}+f_{S}\left( {%
\frac{m_{i}}{k}}\right) kS+f_{C}\left( {\frac{m_{i}}{k}}\right)
C+....\right\} ,  \label{dNdk}
\end{equation}
where area $S=\oint dS$ ($=4\pi R^{2}$ for a sphere) and extrinsic curvature 
$C=\oint \left( {\frac{1}{{R_{1}}}}+{\frac{1}{{R_{2}}}}\right) dS$ ($=8\pi R$
for a sphere). Curvature radii are denoted $R_{1}$ and $R_{2}$. For a
spherical system $R_{1}=R_{2}=R$. The functions $f_{S}$ and $f_{C}$ are
given by \cite{Berger:1986ps,Madsen:1994vp}
\begin{equation}
f_{S}\left( {\frac{m}{k}}\right) =-{\frac{1}{8\pi }}\left\{ 1-\left( {\frac{2%
}{\pi }}\right) \tan ^{-1}{\frac{k}{m}}\right\} ,
\end{equation}
\begin{equation}
f_{C}\left( {\frac{m}{k}}\right) ={\frac{1}{{12\pi ^{2}}}}\left\{ 1-{\frac{3%
}{2}}{\frac{k}{m}}\left( {\frac{\pi }{2}}-\tan ^{-1}{\frac{k}{m}}\right)
\right\} .
\end{equation}

In terms of volume-, surface-, and curvature-densities, $n_{i,V}$, $n_{i,S}$%
, and $n_{i,C}$, the number of quarks of flavor $i$ is 
\begin{equation}
N_i=\int_0^{k_{Fi}}{\frac{{dN_i}}{{dk}}}dk=n_{i,V}V+n_{i,S}S+n_{i,C}C,
\end{equation}
with Fermi momentum $k_{Fi}=(\mu_i^2-m_i^2)^{1/2}$.

The corresponding thermodynamic potentials are related by 
\begin{equation}
\Omega _{i}=\Omega _{i,V}V+\Omega _{i,S}S+\Omega _{i,C}C,
\end{equation}%
where $\partial \Omega _{i}/\partial \mu _{i}=-N_{i}$, and $\partial \Omega
_{i,j}/\partial \mu _{i}=-n_{i,j}$.
For massless quarks $\Omega _{i,S}=n_{i,S}=0$;
$\Omega _{i,C}={\mu _{i}^{2}}/8\pi ^{2}$; $n_{i,C}=-{\ \mu _{i}}/4\pi ^{2}$.

Minimizing the total energy at fixed $N_{i}$ 
gives the pressure equilibrium constraint 
\begin{equation}
B=-\sum_{i}\Omega _{i,V}-{\frac{2}{R}}\sum_{i}\Omega _{i,S}-{\frac{2}{R^{2}}}%
\sum_{i}\Omega _{i,C}.  \label{bag}
\end{equation}

The optimal composition for fixed baryon number, $A$, can be found by
minimizing the energy with respect to $N_i$
giving 
\begin{equation}
0=\sum_i\left( \mu _i+{\frac{{\partial E_{{\rm Coul}}}}{{\partial N_i}}}%
\right) dN_i.  \label{compo}
\end{equation}

For uncharged bulk quark matter one arrives at the usual energy per baryon 
\begin{equation}
\epsilon ^{0}=A^{-1}\sum_{i}N_{i}^{0}\mu _{i}^{0},
\end{equation}%
where superscript $0$ denotes bulk values. The energy minimization, Eq.~(\ref%
{bag}), corresponds to 
\begin{equation}
B=-\sum_{i}\Omega _{i,V}^{0}=\sum_{i}{\frac{{(\mu _{i}^{0})^{4}}}{{4\pi ^{2}}%
}},  \label{bagmin}
\end{equation}
where the last equality assumes massless quarks. In the bulk limit the baryon
number density is given by 
\begin{equation}
n_{A}^{0}={\frac{1}{3}}\sum_{i}{\frac{{(\mu _{i}^{0})^{3}}}{{\pi ^{2}}}},
\end{equation}%
and one may define a bulk radius per baryon as 
\begin{equation}
R^{0}=(3/4\pi n_{A}^{0})^{1/3}.  \label{bulkrad}
\end{equation}

For quark matter composed of massless $u$, $d$, and $s$-quarks, the Coulomb
energy vanishes at equal number densities due to the fact that the sum of
the quark charges is zero. Thus it is energetically most favorable to have
equal chemical potentials for the three flavors. From the equations above
one may derive the following bulk expressions for 3-flavor quark matter: 
\begin{equation}
\mu _i^0=\left( {\frac{{4\pi ^2B}}3}\right) ^{1/4}=1.905B^{1/4}=276.2{\rm MeV%
}B_{145}^{1/4};
\end{equation}
\begin{equation}
n_A^0=(\mu _i^0)^3/\pi ^2=0.700B^{3/4}
\end{equation}
\begin{equation}
R^0=(3/4\pi n_A^0)^{1/3}=0.699B^{-1/4}.
\end{equation}
And the energy per baryon is 
\begin{equation}
\epsilon ^0=3\mu _i^0=5.714B^{1/4},
\end{equation}
in agreement with Eq.\ (\ref{eps3}).

To first order one may regard Coulomb, surface, and curvature energies as
perturbations on top of the bulk solution \cite{Berger:1986ps}. 
In this approach one gets a strangelet mass $M$ ($=E$) 
\begin{eqnarray}
M &=&\epsilon ^{0}A+\sum_{i}\Omega _{i,C}^{0}C^{0}=\epsilon ^{0}A+{\frac{{%
3^{13/12}}B{^{1/4}}A^{1/3}}{{\pi ^{1/6}2^{1/6}}}}  \nonumber \\
&\approx &\left[ 829A{\rm MeV}+351{\rm MeV}A^{1/3}\right] B_{145}^{1/4}.
\label{EoverA}
\end{eqnarray}

For $m_{s}>0$ and $\lambda\equiv m_s/\mu_s$ the energy minimization, 
Eq.~(\ref{bagmin}), changes to 
\begin{equation}
B=\sum_{i=u,d}{\frac{{(\mu _{i}^{0})^{4}}}{{4\pi ^{2}}}}+{\frac{{(\mu
_{s}^{0})^{4}}}{{4\pi ^{2}}}}\left[ (1-\lambda ^{2})^{1/2}(1-{\frac{5}{2}}%
\lambda ^{2})+{\frac{3}{2}}\lambda ^{4}\ln {\frac{{1+(1-\lambda ^{2})^{1/2}}%
}{\lambda }}\right] ,
\end{equation}%
and the baryon number density is now given by 
\begin{equation}
n_{A}^{0}={\frac{1}{3}}\left[ \sum_{i=u,d}{\frac{{(\mu _{i}^{0})^{3}}}{{\pi
^{2}}}}+{\frac{{(\mu _{s}^{0})^{3}}}{{\pi ^{2}}}}(1-\lambda ^{2})^{3/2}%
\right] .
\end{equation}%
A bulk radius per baryon is still defined by Eq.~(\ref{bulkrad}).

In bulk equilibrium the chemical potentials of the three quark flavors are
equal, $\mu _{u}^{0}=\mu _{d}^{0}=\mu _{s}^{0}\equiv \mu ^{0}=\epsilon
^{0}/3 $. Neglecting Coulomb energy one may approximate the mass
of small strangelets as a sum of bulk, surface and curvature terms,
using the chemical potential calculated in bulk: 
\begin{equation}
{M}=\epsilon ^{0}A+\sum_{i}\Omega _{i,S}^{0}S^{0}+\sum_{i}\Omega
_{i,C}^{0}C^{0},
\end{equation}%
where $S^{0}=4\pi (R^{0})^{2}A^{2/3}$ and $C^{0}=8\pi (R^{0})A^{1/3}$.
Masses in MeV for $B^{1/4}=145{\rm MeV}$ are (with $s$-quark mass in MeV
given in parenthesis)

\begin{eqnarray}
M(0) &=&829A+0A^{2/3}+351A^{1/3} \\
M(50) &=&835A+61A^{2/3}+277A^{1/3} \\
M(150) &=&874A+77A^{2/3}+232A^{1/3} \\
M(200) &=&896A+53A^{2/3}+242A^{1/3} \\
M(250) &=&911A+22A^{2/3}+266A^{1/3} \\
M(300) &=&917A+0.3A^{2/3}+295A^{1/3} \\
M(350) &=&917A+0A^{2/3}+296A^{1/3}  \label{udmass}
\end{eqnarray}

The lack of a significant Coulomb energy is due to the fortuitous
cancellation of charge $+2e/3$ up quarks and charge $-e/3$ down and strange
quarks in strange quark matter with equal numbers of the three quark
flavors. Because of the non-zero $s$-quark mass the cancellation is not
perfect. Typical strangelets have slightly fewer strange quarks compared to
up and down, and therefore the net charge is slightly positive. A typical
model result (to be compared to $Z\approx 0.5A$ for nuclei) is
\cite{Heiselberg:1993dc}

\begin{align}
Z& =0.1\left( \frac{m_{s}}{150{\rm MeV}}\right) ^{2}A \\
Z& =8\left( \frac{m_{s}}{150{\rm MeV}}\right) ^{2}A^{1/3}
\end{align}%
for $A\ll 700$ and $A\gg 700$ respectively (the slower growth for higher $A$
is a consequence of charge screening).

Thus a unique experimental signature of strangelets is an unusually high
mass-to-charge ratio compared to nuclei.

Cooper pairing involves quarks with equal (but opposite) momenta, so the
natural ground state of a color-flavor locked system has equal Fermi momenta
for up, down, and strange quarks and therefore equal number
densities. Thus the total net quark charge is zero for a bulk system
\cite{Rajagopal:2000ff}.
A finite strangelet has a surface suppression of massive
strange quarks relative to the almost massless ups and downs (massive
particle wave functions are suppressed at a surface), so the total
charge of a color-flavor locked strangelet is positive and proportional to
the surface area \cite{Madsen:2001fu,Madsen:2000kb}:

\begin{equation}
Z=0.3\left( \frac{m_{s}}{150{\rm MeV}}\right) A^{2/3}.
\end{equation}

This phenomenon persists even for very large bags, such as strange stars, so
color-flavor locked strange stars also have a positive quark charge
\cite{Madsen:2001fu,Usov:2004iz,Stejner:2005mw}.

Writing $M=\epsilon ^0A+c_{{\rm surf}}A^{2/3}+c_{{\rm curv}}A^{1/3}$,
with $c_{{\rm surf}}\approx 100$MeV and $c_{{\rm curv}}\approx 300$MeV, the
stability condition $M<Am_n$ may be written as $A>A_{{\rm min}}^{{\rm abs}}$%
, where 
\begin{equation}
A_{{\rm min}}^{{\rm abs}}=\left( {\frac{{c_{{\rm surf}}+[c_{{\rm surf}%
}^2+4c_{{\rm curv}}(m_n-\epsilon ^0)]^{1/2}}}{{2(m_n-\epsilon ^0)}}}\right)
^3.
\end{equation}
Stability at baryon number 30 requires a bulk binding energy in excess of 65
MeV, which is barely within reach in unpaired strange quark matter 
for $m_s>100$MeV if, at the same time, $ud$-quark matter shall be unstable.
For color-flavor locked strangelets, stability is more likely.
Long-lived metastability with respect to neutron emission is possible if $%
dE_{{\rm curv}}/dA+dE_{{\rm surf}}/dA<m_{n}-\epsilon ^{0}$, or $A$ larger
than 
\begin{equation}
A_{{\rm \min }}^{{\rm meta}}=\left( {\frac{{c_{{\rm surf}}+[c_{{\rm surf}%
}^{2}+3c_{{\rm curv}}(m_{n}-\epsilon ^{0})]^{1/2}}}{{3(m_{n}-\epsilon ^{0})}}%
}\right) ^{3}.
\end{equation}
To have $A_{{\rm min}}^{{\rm meta}}<30$ requires $m_{n}-\epsilon ^{0}>30$ MeV.

Shell effects can have a stabilizing effect not taken into account in
the liquid drop model approach above. As stressed by
Gilson and Jaffe \cite{Gilson:1993zs} 
the fact that the slope of $E/A$ versus $A$
becomes very steep near magic numbers can lead to strangelets that are
metastable (stable against single baryon emission) even for $\epsilon
^{0}>930$MeV.

\section{The AMS and LSSS experiments}

Strangelet (meta)stability is a theoretical possibility as demonstrated
above, but the existence of small baryon
number strangelets is ultimately an experimental issue.

Several experiments have searched for strangelets in cosmic rays. While some
interesting events have been found that are consistent with the predictions
for strangelets, none of these have been claimed as real discoveries.
Whether interpreted as flux limits or as detections these results are
consistent with the flux predictions from strange star collisions of a
few strangelets per year per square meter per steradian given in 
Ref.~\citen{Madsen:2004vw} (see also 
Refs.~\citen{Madsen:1989pg,Friedman:1990qz,
Benvenuto:1989hw,Medina-Tanco:1996je,Cheng:2006ak}). More specifically
the integrated flux of mass $A$, charge $Z$ strangelets reaching the
inner Solar System was predicted to be\cite{Madsen:2004vw}
\begin{equation}
F=2\times 10^5 A^{-1.067}Z^{-0.6}
\end{equation}
per year per square meter per steradian under the conservative assumption of
$10^{-10}$ solar masses per year of strangelets being released in our
Galaxy from binary strange star collisions. The major unknown is the
mass spectrum of strangelets released, but it seems plausible within
quite different scenarios that some strangelets have sufficiently low
mass to be detectable at a reasonably high rate 
\cite{Madsen:2001bw,Jaikumar:2005ne}.

Two experiments that are
currently underway will reach sensitivities that would provide a
definitive strangelet detection or rule out a significant part of
parameter space.

{\bf AMS-02: }The Alpha Magnetic Spectrometer (AMS) is a space-based
particle physics experiment involving several hundred physicists from more
than 50 institutions in 16 countries, led by Samuel Ting of
MIT. A prototype (AMS-01) flew in June 1998 aboard the Space Shuttle
Discovery \cite{Aguilar:2002ad}, and AMS-02 is currently scheduled to fly to
the International Space Station (ISS) in 2008. Once on the ISS AMS-02 will
remain active for at least three years. Equipped with a superconducting
magnet, time-of-flight detectors, trackers, calorimeter, a ring imaging
Cerenkov counter, etcetera, AMS-02 will provide data with
unprecedented accuracy on cosmic ray electrons, positrons, protons, nuclei,
anti-nuclei and gammas in the GV-TV range and probe issues such as
antimatter, dark matter, cosmic ray formation and propagation. In addition
it will be uniquely suited to discover strangelets characterized by extreme
rigidities for a given velocity compared to nuclei 
\cite{Sandweiss:2004bu,Finch:2006pq}.
AMS-02 will have excellent charge resolution up to $Z\approx26$, 
and should
be able to probe a large mass range for strangelets. A reanalysis of data
from the 1998 AMS-01 mission has given hints of some interesting events,
such as one with $Z=2,A=16$ \cite{Choutko:2003} and another with $Z=8,$ but
with the larger AMS-02 detector running for 3 years or more, real statistics
is achievable.

{\bf LSSS: }The Lunar Soil Strangelet Search (LSSS) is a search for $Z=8$
strangelets using the tandem accelerator at the Wright Nuclear Structure
Laboratory at Yale. The experiment involves a dozen people from Yale, MIT,
and \AA rhus, led by Jack Sandweiss of Yale and
studies a sample of 15 grams of lunar
soil from Apollo 11. It will reach a sensitivity of $10^{-17}$ over a wide
mass range \cite{Han:2006,Monreal:2005dg}, 
sufficient to provide detection according to the estimates
in Ref.~\citen{Madsen:2004vw} if strangelets have been trapped in 
the lunar surface layer.
In contrast to the deep oceanic and geological mixing on Earth, the
effective mixing depth of the lunar surface is only around one meter,
determined largely by micrometeorite impacts. Therefore the surface
layer has an effective cosmic ray exposure time of around 500 million years.
Combined with the lack of a shielding magnetosphere, this results in an
expected strangelet concentration in lunar soil which is four orders of
magnitude larger than the concentration on Earth, where the search for
strangelets so far has resulted in upper limits only.

\section{Strangelets as ultra-high energy cosmic rays}

Strangelets may even provide an explanation for one of the most
interesting mysteries in cosmic ray physics:
The existence of cosmic rays with energies well beyond $10^{19}{\rm eV}$,
with measured energies as high as $3\times10^{20}{\rm eV}$
\cite{Greisen:1966jv,Zatsepin:1966jv}. It is almost impossible to find a
mechanism to accelerate cosmic rays to these energies. Furthermore
ultra-high energy cosmic rays lose energy in interactions with cosmic
microwave background photons, and only cosmic rays from nearby
(unidentified) sources would reach us with the energies measured.
Strangelets circumvent both problems, and therefore provide a possible
mechanism for cosmic rays beyond the socalled Greisen-Zatsepin-Kuzmin
(GZK) cutoff \cite{Madsen:2002iw}:

{\bf Acceleration: }All astrophysical \textquotedblleft
accelerators\textquotedblright\ involve electromagnetic fields, and the
maximal energy of a charged particle is proportional to its charge. The
charge of massive strangelets has no upper bound in contrast to nuclei, so
highly charged strangelets are capable of reaching energies much higher than
those of cosmic ray protons or nuclei using the same \textquotedblleft
accelerator\textquotedblright\ \cite{Madsen:2002iw}.

{\bf The GZK-cutoff} is a consequence of ultrarelativistic cosmic rays
hitting a $2.7{\rm K}$ background photon with a Lorentz-factor $\gamma$
large enough to boost the $7\times10^{-4}\,{\rm eV}$ photon to energies
beyond the threshold of energy loss processes, such as photo-pion production
or photo-disintegration. The threshold for such a process has a fixed
energy, $E_{{\rm Thr}}$, in the frame of the cosmic ray, e.g., $E_{{\rm Thr}%
}\approx10{\rm MeV}$ for photo-disintegration of a nucleus or a strangelet,
corresponding to $\gamma_{{\rm Thr}}=E_{{\rm Thr}}/E_{2.7{\rm K}%
}\approx10^{10},$ or a cosmic ray total energy 
\begin{equation}
E_{{\rm Total}}=\gamma_{{\rm Thr}}Am_{0}c^{2}\approx10^{19}A~{\rm eV.}
\end{equation}
Since strangelets can have much higher $A$-values than nuclei, this pushes
the GZK-cutoff energy well beyond the current observational limits for
ultra-high energy cosmic rays \cite{Madsen:2002iw,Rybczynski:2001bw}.

\section{Conclusion}

Strange quark matter may be absolutely stable in bulk, and smaller lumps
(strangelets) can be stable down to some small limiting baryon number.
Strangelets may
form in a first-order cosmological quark-hadron phase transition (unlikely),
or in processes related to ccompact stars (more likely). Flux estimates for
lumps reaching our neighborhood of the Galaxy as cosmic rays are in a range
that makes it realistic to either detect them in upcoming experiments like
AMS-02 or LSSS, or place significant limits on the existence of stable strange
quark matter. Should strangelets be discovered it would have profound
consequences for our understanding of the strong interactions and for a
variety of astrophysical phenomena.

\section*{Acknowledgments}

This work was supported by the Danish Natural Science Research Council.

\end{document}